\documentclass[12pt]{article}
\usepackage{amsmath,amssymb,amsfonts,amsthm}

\title{Dynamics of entropy and nonclassical properties of the state of a $\Lambda$-type three-level atom
 interacting with a single-mode cavity field  with intensity-dependent coupling in a Kerr medium}

\author{M J Faghihi$^{1}$ and M K Tavassoly$^{1,2}$
\\
\footnotesize{$^{1}$ Atomic and Molecular Group, Faculty of Physics, Yazd University, Yazd, Iran} \\
\footnotesize{$^{2}$ Research Group of Optics and Photonics, Yazd University, Yazd, Iran}
\\ \footnotesize{E-mail: mktavassoly@yazd.ac.ir}}

\begin{document}
\maketitle


 \begin{abstract}
 In this paper, we study the interaction between a three-level atom and a quantized single-mode field with $` `$intensity-dependent coupling$"$ in a $` `$Kerr medium$"$.
 The three-level atom is considered to be in a $\Lambda$-type configuration.
 Under particular initial conditions, which may be prepared for the atom and the field, the dynamical state vector of the entire system
 will be explicitly obtained,  for arbitrary nonlinearity function $f(n)$ associated to any physical system.
 Then, after evaluating the variation of the field entropy against time, we will investigate the
 quantum statistics as well as some of the nonclassical properties of the introduced state. During our calculations we investigate the effects of intensity-dependent
 coupling, Kerr medium and detuning parameters on the depth and domain of the nonclassicality features of the atom-field state vector. Finally, we compare our obtained results with those of $V$-type three-level atoms.
 \end{abstract}

 {\bf Pacs:} 42.50.Ct, 42.50.Dv, 42.50.Ar, 42.50.-p

 {\bf Keywords:}
 Atom-field interaction; intensity-dependent coupling; Kerr medium;  $\Lambda$-type three-level atom;
 Nonclassical state.

 %
 \section{Introduction}\label{sec-intro}
 %
 As is well-known, the full quantum mechanical atom-field interaction can predict new aspects of quantum nature of the field
 as well as the atom. One of the most common models in quantum optics, which is usually used for the description of the atom-field interaction, is the Jaynes-Cummings model (JCM). This model is basically a full quantum mechanical theory that gives a solution for the
 interaction between a two-level atom and a single-mode field \cite{JCM, cum}. Many kinds of generalizations have been proposed to modify the JCM, through which the frequent usefulness of such a model has been shown in various physical studies. Indeed, the JCM and its generalizations can predict new physical results. JCM with intensity-dependent coupling suggested by Buck and Sukumar \cite{suk1,suk2} describes the dependence of atom-field coupling on the light intensity. Bu$\check{\mathrm{z}}$ek \cite{buzek} demonstrated that in the intensity-dependent coupling JCM, the exact periodicity of the physical quantities, particularly the atomic population inversion and the squeezing, may be observed if one considers the interaction between a two-level atom and a single-mode field. Moreover, the latter observations can be destroyed if the radiation field interacts with a system of more than one two-level atom, or when more than two levels of a single-atom are taken into account \cite{koroli}. An-fu and Zhi-wei \cite{an-fu} found that the phase distribution depends on the coherent field intensity and the detuning parameter. Fang {\it et al} examined the properties of entropy and phase of the field in two-photon JCM in the presence of nonlinear interaction of a Kerr medium with the field mode \cite{mao}. The effect of virtual-photon fields in the JCM with the counter-rotating wave terms has been investigated by Xie  {\it et al} \cite{rui}. Kazakov \cite{kazakov} discussed the independent interaction of a two-level atom with two-modes quantized field, according to the modified JCM. Marchiolli \cite{marcelo} examined,  in the framework of the non-resonant JCM, the nonclassical properties of a  two-level atom interacting with a single-mode cavity field which is prepared in a finite and discrete harmonic oscillator-like coherent state. Crnugelj {\it et al} \cite{crn} studied the time evolution and squeezing properties of a deformed JCM, which corresponds to the usual model with intensity-dependent coupling, controlled by two additional parameters may be determined by experiment.
 The collapse and revival in the JCM have  been analyzed analytically by Feranchuk {\it et al} \cite{feran}. Recently, Koroli {\it et al}
 \cite{Koroli} studied the interaction of an equidistant three-level
 atom (ion), whose dipole moment matrix transition elements between the adjacent atomic levels
 are different from the GP nonlinear coherent state of $\mathrm{SU}(1,1)$ group. More recently, a nonlinear interaction between a two-level atom and a single-mode field is considered
 in the frame of JCM by Cordero {\it et al}  \cite{sergio} and the complete revivals are established for appropriate coupling.
 Evolution of a two-level atom in a strong resonant quantum field beyond the rotating wave approximation (RWA)
 has been considered in \cite{feran2}
 and recently, the nonlinear interaction of an equidistant three-level atom and a single-mode cavity field,
 that is initially prepared in a generalized coherent state, has been studied by one of us \cite{yadollahi}.
 Also, before these, three-level atoms have been studied in somewhat different ways. For instance, Radmore and Knight \cite{radmore} discussed the time evolution of a three-level system in both ladder and $\Lambda$-configurations driven by two different fields of arbitrary amplitudes and detunings using dressed states.\\
 Now, in this paper, we study a $\Lambda$-type three-level atom
 interacting with a single-mode field, regarding the JCM with intensity-dependent coupling between the atom  and the field (nonlinear JCM) surrounded by a Kerr medium.
 Recently, such an interaction for a
 three-level atom in $V$-configuration has been studied in
 \cite{zait1}. A special case of \cite{zait1}, in which the Kerr
 medium is neglected and the intensity-dependent function is
 considered to be $f(n)=\sqrt{n}$, has been discussed in \cite{huang}.
 Moreover, since $V$- and $\Lambda$-configurations are adequately
 distinguishable due to their different structures and properties, investigating
 this type of interaction in a $\Lambda$-type three-level atom,
 from which we will obtain new results, seems to be useful.
 Anyway, along this perspective, our further motivation of this paper is to investigate
 the effects of Kerr medium, intensity-dependent coupling and
 detuning parameters on the physical properties such as field
 entropy, Mandel's $Q$ parameter, normal and higher-order squeezing and quasiprobability $Q$-distribution function.\\
 The paper is organized as follows. In the next section, we attempt to find the explicit form of the state vector
 of the atom-field system  using the nonlinear JCM. In section 3, the entropy evolution of the field is evaluated. Then, we pay attention to the quantum statistics of the obtained states by considering the Mandel $Q$ parameter in section 4.
 Normal squeezing, amplitude-squared squeezing and amplitude-cubed squeezing of the obtained state are studied in section 5.
 Section 6 deals with the quasiprobability $Q$-distribution function. Section 7 contains a summary and concluding remarks.
 %
 %

 \section{Introducing the state vector of the system}
 %
 In quantum mechanics, the most important step in studying any
 physical system is the construction of an appropriate Hamiltonian
 of the system. This goal is achieved by an exact view on the
 existing interactions between subsystems. Then, by solving the
 time-dependent Schr\"{o}dinger equation, one may find the
 dynamical state of the system under study. Possible information arises
 from the wave function of the system.
 Let us consider a
 model in which the single-mode electromagnetic field which oscillates
 with frequency $\Omega$ in an optical cavity involving a Kerr medium interacts with a $\Lambda$-type three-level atom.
 The three levels of the atom are indicated by $|j\rangle$
 with energies $\omega_{j}$ where $j=1,2,3$ (see figure 1). In this type of atoms, the transitions $|1\rangle\rightarrow|2\rangle$ and
 $|1\rangle\rightarrow|3\rangle$ are allowed and the transition
 $|2\rangle\rightarrow|3\rangle$ is forbidden in the electric-dipole approximation \cite{zubairy}. The
 Hamiltonian for this system in the RWA can be written as ($\hbar=1=c$)
 \begin{eqnarray}\label{hamiltonih}
 H =H_{0}+H_{1},
 \end{eqnarray}
 where
 \begin{eqnarray}\label{hamiltonih0}
 H_{0}&=& \sum_{j=1}^{3} \omega_{j}\sigma_{jj}+\Omega a^{\dag} a,
 \end{eqnarray}
 \begin{eqnarray}\label{hamiltonih1}
 H_{1}=\chi a^{\dag 2} a^2+\lambda_{1}(R\sigma_{13}+\sigma_{31}R^{\dag})
 + \lambda_{2}(R\sigma_{12}+\sigma_{21}R^{\dag}),
 \end{eqnarray}
 where $\sigma_{ij}$ denotes the lowering and raising operators between $|i\rangle$ and
 $|j\rangle$ defined by $\sigma_{ij}=|i\rangle \langle j| (i,j=1,2,3),a$ and $a^{\dag}$ are respectively bosonic annihilation and
 creation operators of the field, $\chi$ denotes the dispersive part of the third-order nonlinearity of the Kerr medium and the constants
 $\lambda_{1},\lambda_{2}$ determine the atom-field coupling. A deep insight in the form of the introduced $H_{1}$ in (\ref{hamiltonih1})
 demonstrates that this Hamiltonian is constructed by changing $\lambda_{i}$ to $\lambda_{i}f(n), i=1,2$, when compared with the standard JCM. Henceforth,
 sometimes it is called $` `$nonlinear JCM$"$ \cite{buzek}. The operators $R$ and $R^{\dag}$ are respectively the nonlinear ($f$-deformed)
 annihilation and creation operators, which satisfy the following communication relations:
 \begin{eqnarray}\label{vrrd2}
 \left[R, R^{\dag}\right] &=& (n+1)f^2(n+1)-n f^2(n),\;\;\;\left[R,n\right] = R, \;\;\; \left[R^{\dag},n\right]=-R^{\dag},
 \end{eqnarray}
 where $n=a^{\dag}a$, $R=af(n)$, $R^{\dag}=f(n)a^{\dag}$ and $f(n)$ is a Hermitian operator-valued function responsible for the intensity-dependent atom-field coupling. In order to rearrangement the Hamiltonian in (\ref{hamiltonih0}), we use the Heisenberg equation of motion and  the constants of motion will be obtained.
 These constants contain atomic probability and excitation number which can be expressed by the following relations \cite{knight}:
 \begin{eqnarray}\label{constant}
 P_{A}&=&\sum_{i=1}^{3}\sigma_{ii}=I, \nonumber \\
 \mathbb{N}&=&a^{\dag}a+\sigma_{11},
 \end{eqnarray}
 where $I$ is the unity operator. Hence, the free part of the Hamiltonian in (\ref{hamiltonih0}) can be rewritten as
 \begin{eqnarray}
 H_{0}=\omega_{3}I+\Omega \;\mathbb{N}-(\Delta_{3}-\Delta_{2})\sigma_{22}-\Delta_{3}\sigma_{11},
 \end{eqnarray}
 where the detuning parameters $\Delta_{2}$ and $\Delta_{3}$ are given by
 \begin{eqnarray}\label{cavp}
 \Delta_{2}=\omega_{2}-\omega_{1}+\Omega, \;\;\;\;\;  \Delta_{3}=\omega_{3}-\omega_{1}+\Omega.
 \end{eqnarray}
 The wave function $|\psi(t)\rangle$ at any time $t$ may be written in the following form
 \begin{eqnarray}\label{say}
 |\psi(t)\rangle&=&\sum_{n=\circ}^{\infty}q_{n}\Big[ A(n,t) e^{-i\gamma_{1} t}|1,n\rangle+B(n+1,t)e^{-i\gamma_{2} t}|2,n+1\rangle \nonumber \\
 &+&  C(n+1,t)e^{-i\gamma_{3} t}|3,n+1\rangle \Big],
 \end{eqnarray}
 where $q_{n}$ describes the amplitude of the initial field state, $A,B$ and $C$ are the atomic probability amplitudes which have to be determined and
 \begin{eqnarray}
 \gamma_{1}&=&\omega_{1}+n\Omega, \nonumber \\
 \gamma_{2}&=&\omega_{2}+(n+1)\Omega, \nonumber \\
 \gamma_{3}&=&\omega_{3}+(n+1)\Omega.
 \end{eqnarray}
 At this stage, the exact values of atomic probability amplitudes that determine the explicit form of the wave function should be found. For this purpose, we insert  the assumed wave function (\ref{say}) into the time-dependent Schr\"{o}dinger equation considering the Hamiltonian in (\ref{hamiltonih}). Farther calculations lead to the following coupled differential equations for the probability amplitudes
 \begin{eqnarray}\label{coupling}
 i\;\dot{A}&=& V_{1}A+ f_{1}Ce^{-i\Delta_{3}t}+f_{2}Be^{-i\Delta_{2}t}, \nonumber \\
 i\;\dot{B}&=&V_{2}B+f_{2}Ae^{i\Delta_{2}t}, \nonumber \\
 i\;\dot{C}&=&V_{2}C+f_{1}Ae^{i\Delta_{3}t},
 \end{eqnarray}
 where the dot signs refer to the time differentiation and we have set
 \begin{eqnarray}\label{vdefinition}
 f_{1}&\dot{=}&\lambda_{1}\;\sqrt{n+1}\;f(n+1), \nonumber \\
 f_{2}&\dot{=}&\lambda_{2}\;\sqrt{n+1}\;f(n+1), \nonumber \\
 V_{1}&\dot{=}&\chi\; n(n-1), \nonumber \\
 V_{2}&\dot{=}&\chi \;n(n+1).
 \end{eqnarray}
 By assuming  $B=e^{i\mu t}$ and inserting it into the equations expressed in ({\ref{coupling}}), one arrives at
 \begin{eqnarray}\label{mu}
 \mu ^{3}+x_{1} \mu ^{2}+x_{2} \mu + x_{3}=0,
 \end{eqnarray}
 where
 \begin{eqnarray}\label{x123}
 x_{1}&\dot{=}&2V_{2}+V_{1}+\Delta_{3}-\Delta_{2}, \nonumber \\
 x_{2}&\dot{=}& \Delta_{2}\left(\Delta_{2}-\Delta_{3}-2V_{2}-V_{1}\right)+V_{2}\left(2V_{1}+V_{2}+\Delta_{3}\right)
 +V_{1}\Delta_{3}-f_{1}^{2}-f_{2}^{2},  \nonumber \\
 x_{3}&\dot{=}&\left[(V_{2}+\Delta_{3}-\Delta_{2})(V_{1}-\Delta_{2})-f_{1}^{2}-f_{2}^{2}\right]V_{2}
 -f_{2}^{2}(\Delta_{3}-\Delta_{2}).
 \end{eqnarray}
 It is clear that Eq. (\ref{mu}) has generally three different roots. Hence, $B$ can be implied as a linear combination of $e^{i\mu_{j}t}$ as follows:
 \begin{eqnarray}\label{b}
 B=\sum_{j=1}^{3} \tilde{b_{j}}e^{i\mu_{j} t},
 \end{eqnarray}
 where $\tilde{b_{j}}=f_{2}b_{j}$. With the help of Kardan instruction \cite{kardan}, the general solution of Eq. (\ref{mu}) is given by
 \begin{eqnarray}\label{vkardan}
 \mu_{j}&=&-\frac{1}{3}x_{1}+\frac{2}{3}\sqrt{x_{1}^{2}-3x_{2}}\cos\left[ \theta+\frac{2}{3}(j-1)\pi \right],\;\;\;\;\;j=1,2,3, \nonumber \\
 \theta &=& \frac{1}{3}\cos^{-1}\left[ \frac{9x_{1}x_{2}-2x_{1}^{3}-27x_{3}}{2(x_{1}^{2}-3x_{2})^{3/2}}\right].
 \end{eqnarray}
 Finally, by replacing Eq. (\ref{b}) into the coupled differential equations in (\ref{coupling})
 and after some lengthy but straightforward manipulations, we find the probability amplitudes in the form
 \begin{eqnarray}\label{abc}
 A(n,t)&=&-e^{-i\Delta_{2}t}\sum_{j=1}^{3}(\mu_{j}+V_{2})b_{j}e^{i\mu_{j}t}, \nonumber \\
 B(n+1,t)&=&\sum_{j=1}^{3}f_{2}\;b_{j} e^{i\mu_{j} t}, \nonumber \\
 C(n+1,t)&=&\frac{1}{f_{1}}e^{i(\Delta_{3}-\Delta_{2})t}\sum_{j=1}^{3}\Big[(\mu_{j}+V_{2})
 (\mu_{j}+V_{1}-\Delta_{2})-f_{2}^{2}\Big]b_{j}e^{i\mu_{j}t},
 \end{eqnarray}
 where $b_{j}$ can be determined by the initial conditions for the probability amplitudes. Now, let us consider the atom to be initially in the excited state, i.e. $A(0)=1$, $B(0)=C(0)=0$; then using (\ref{abc}) the following relations may be found:
 \begin{eqnarray}\label{b123}
 b_{j}=\frac{\mu_{k}+\mu_{l}+V_{1}+V_{2}-\Delta_{2}}{\mu _{jk} \mu _{jl}},\;\;\;\;\;\;j\neq k\neq l=1,2,3,
 \end{eqnarray}
 where $ \mu _{jk}=\mu_{j}-\mu_{k} $. In conclusion, as is seen, the wave function $|\psi(t)\rangle$ introduced in (\ref{say}) is explicitly obtained.
 Therefore, we are now able to study the nonclassical properties of the state of the atom-field system,
 of course after specifying the initial state of the field.
 %
 \section{Entropy evolution of the field}
 %
 %
 In recent years, various research works have been progressed in quantum entanglement, which is one of the main parts for the execution
 of quantum information processing devices \cite{qipd}. From our obtained results in the previous section, it may be understood that quantum dynamics associated with the presented atom-field quantum system leads to the entanglement between the atom and the field. On the other hand, the entropy of the field is a criterion which implies the strength of entanglement. The higher (lower) the entropy, the greater (smaller) the degree of entanglement. Hence, we are reasonably interested in the investigation of the time evolution of the entropy of our obtained state. To study the dynamics of the entanglement, one must choose an entanglement measure. For the present case, we use the linear entropy or von Neumann reduced entropy \cite{pk1}. However, before investigating the field entropy, the important theorem of Araki and Leib is worth recalling \cite{araki}. According to this theorem, for any two-components of quantum systems (for instance the one under consideration), the entropies are limited by the following triangle inequality:
 \begin{eqnarray}\label{valen}
 |S_{A}-S_{F}|\leq S \leq |S_{A}+S_{F}|,
 \end{eqnarray}
 where here the subscripts $` `$A$"$ and $` `$F$"$ refer to the atom and the field, respectively. The total entropy of the atom-field
 system is denoted by S. If at the initial time, the field and the atom are in pure states, the total entropy of the system is zero and remains constant. So, assuming initially, $S=0$ (if the system is prepared in a pure state), at any time $t>0$, the entropy of the field is equal to the atomic entropy \cite{phoenix}. Therefore, instead of the evaluation of the field entropy, we can obtain the entropy of the atom. The entropy of the atom (field) according to the von Neumann entropy is defined through the corresponding reduced density operator by
 \begin{eqnarray}\label{ventd}
 S_{A(F)}(t)=-\mathrm{Tr}_{A(F)} \left(\rho_{A(F)} \ln \rho_{A(F)} \right).
 \end{eqnarray}
 The reduced density matrix of the atom required for evaluating (\ref{ventd}) is given by
 \begin{eqnarray}\label{vrdma}
 \rho_{A}(t)&=&\mathrm{Tr}_{F}\left(  |\psi \rangle \langle \psi |   \right) \nonumber \\
 &=&\left(
  \begin{array}{ccc}
    \rho_{11} & \rho_{12} & \rho_{13} \\
    \rho_{21} & \rho_{22} & \rho_{23} \\
    \rho_{31} & \rho_{32} & \rho_{33} \\
  \end{array}
 \right).
 \end{eqnarray}
 The matrix elements in (\ref{vrdma}) are given, for instance, by
 \begin{eqnarray}\label{rho}
 \rho_{11}&=&\sum_{n=0}^{\infty}P_{n}A(n,t)A^{*}(n,t), \nonumber \\
 \rho_{12}&=&\sum_{n=0}^{\infty}P_{n}A(n,t)B^{*}(n+1,t) e^{i \Delta_{2} t}, \nonumber \\
 \rho_{13}&=&\sum_{n=0}^{\infty}P_{n}A(n,t)C^{*}(n+1,t)e^{i \Delta_{3} t},... \nonumber
 \end{eqnarray}
 where in all of the above relations, $P_{n}=|q_{n}|^{2}$ is the distribution of the initial radiation field, and $A,B$ and $C$ are the
 atomic probability amplitudes derived in (\ref{abc}). Hence, the entropy of the field or atom can be obtained by the following
 relation \cite{pk2, pk3}:
 \begin{eqnarray}\label{sff}
 S_{F}(t)=S_{A}(t)=-\sum_{j=1}^{3}\xi_{j} \ln \xi_{j},
 \end{eqnarray}
 where $\xi_{j}$, the eigenvalues of the reduced atomic density matrix in (\ref{vrdma}), read as
 \begin{eqnarray}\label{ventkardan}
 \xi_{j}&=&-\frac{1}{3}\alpha_{1}+\frac{2}{3}\sqrt{\alpha_{1}^{2}-3\alpha_{2}}\cos\left[\beta+\frac{2}{3}(j-1)\pi \right],\;\;\;\;\;j=1,2,3, \nonumber \\
 \beta &=& \frac{1}{3}\cos^{-1}\left[ \frac{9\alpha_{1}\alpha_{2}-2\alpha_{1}^{3}-27\alpha_{3}}{2(\alpha_{1}^{2}-3\alpha_{2})^{3/2}}\right],
 \end{eqnarray}
 with
 \begin{eqnarray}\label{vzal}
 \alpha_{1}&\dot{=} &-\rho_{11}-\rho_{22}-\rho_{33}, \nonumber \\
 \alpha_{2}&\dot{=}&\rho_{11}\rho_{22}+\rho_{22}\rho_{33}+\rho_{33}\rho_{11}-\rho_{12}\rho_{21}-\rho_{23}\rho_{32}-\rho_{31}\rho_{13}, \nonumber \\
 \alpha_{3}&\dot{=}& -\rho_{11}\rho_{22}\rho_{33}-\rho_{12}\rho_{23}\rho_{31}-\rho_{13}\rho_{32}\rho_{21} \nonumber \\
 &+&\rho_{11}\rho_{23}\rho_{32}+\rho_{22}\rho_{31}\rho_{13}+\rho_{33}\rho_{12}\rho_{21}.
 \end{eqnarray}
 Studying the dynamical behaviour of the entropy of the system under consideration leads us to obtain a correlation between the atom and the field. Equation (\ref{sff}) determines the variation of the entropy of the atom or the field with time. In addition, we note that by this equation the degree of entanglement between the atom and field is also determined, i.e. the subsystems are disentangled (the system of atom-field is separable) if equation (\ref{sff}) tends to zero.\\
 For simplicity and without loss of generality, we consider $\lambda_{1}=\lambda_{2}=\lambda$ in all of our numerical calculations in the remainder of the paper.
 Henceforth, we can plot all required quantities as a function of scaled time $\tau=\lambda t$. Also, the field is considered to be initially in a coherent state:
 \begin{eqnarray}\label{vqf}
 |\alpha \rangle&=&\exp\left(-\frac{|\alpha|^{2}}{2}\right)\sum_{n=\circ}^{\infty}\frac{\alpha^{n}}{\sqrt{n!}}\;|n \rangle.
 \end{eqnarray}
 So,
 \begin{eqnarray}
 P_{n}=|q_{n}|^{2}=\exp(-|\alpha|^{2})\frac{|\alpha|^{2n}}{n!},
 \end{eqnarray}
 where $|\alpha|^{2}$ is exactly the mean photon number (intensity of light) of the initial coherent field.
 We present the numerical results for the cases $f(n)=1$ and $f(n)=1/\sqrt{n}$, which have been obtained by Man'ko {\it et al} \cite{manko} (where the corresponding coherent states have been named by Sudarshan as harmonious states \cite{harmonios}).
 At this point, we would like to mention that we have used exactly the same nonlinearity function and the same parameters of \cite{zait1}
 in all of our further numerical results, so as to make a better comparison with the $V$-type configuration \cite{zait1}, which
 will be offered in the continuation of the paper.\\
 Figure 2 shows the evolution of the field entropy against the scaled time $\tau$ for initial mean number of photons fixed at
 $|\alpha|^{2} = 10$.  The left plots concern with the absence of the intensity-dependent coupling, i.e. $f (n) = 1$
 and in the right plots  the intensity-dependent coupling  with $f (n) = 1/\sqrt{n}$ is considered.
 In figure 2(a) the Kerr medium is eliminated  ($\chi = 0$) and the exact resonant case is considered ($\Delta_{2} = \Delta_{3} =0$).
 figure 2(b) shows the effect of the Kerr medium ($\chi/\lambda = 0.4$) in exact resonance condition. The effect of the detuning
 parameters ($\Delta_{2}/\lambda =7,  \Delta_{3}/\lambda =15$) in the absence of the Kerr medium ($\chi = 0$) has been shown in figure 2(c).
 As seen from the left plot of figure 2(a), whereas the intensity-dependent coupling and the Kerr effect are disregarded,
 a chaotic behaviour for the time evolution of the field entropy is revealed in the resonant case. By including the intensity-dependent coupling (right plot of figure 2(a)) a regular oscillatory behaviour is observed. Instead, in both plots of figure 2(b), the behaviour of entropy of the field is clearly chaotic. Furthermore, comparing the right plots of figures 2(a) and 2(b) indicates that the Kerr effect in the presence of the intensity-dependent coupling reduces the maximum value of the quantum field entropy, while it is seen that the minimum value of the entropy has been increased. The effect of the detuning parameters can be seen from figure 2(c).
 One can observe that while in both plots of figure 2(c) the time evolution of the field entropy has a chaotic temporal behaviour with  rapidly changes against scaled time, the presence of intensity-dependent coupling
 (right  plot of figure 2(c)) causes the maxima of oscillations to be lowered by a relative amount of nearly  $0.40$.
 Also, comparing the right plots of figures 2(a) and 2(c) shows that the detuning parameters descend the amount of the field entropy.
 Both of these figures have a periodic behaviour against the scaled time $\tau$. It appears that the presence of the detuning effect causes an increase in the period of time for the field entropy.
 We conclude this section with comparing our presented results with \cite{zait1}. From figure 2 it is clear that for all cases,
 the quantum field entropy and subsequently the degree of entanglement between atom and field
 for the $\Lambda$-type three-level atom is greater than $V$-type configuration.
 We note that the value of zero in these plots expresses that the atom and the field are disentangled.
 In fact, at times when the entropy becomes zero, the atom is in its pure states.
 %
 %
 \section{Photon statistics: the Mandel parameter}
 %
 To study the statistical properties of the system, Mandel parameter is usually a helpful quantity. This parameter has been defined as follows \cite{mandel1}:
 \begin{eqnarray}\label{mandel1}
 Q = \frac{\langle (\Delta n)^2\rangle - \langle n \rangle}{\langle n \rangle},
 \end{eqnarray}
 where $(\Delta n)^{2}=\langle n^{2} \rangle-\langle n \rangle^{2}$. When $-1\leq Q <0 \;(Q>0)$, the statistics is sub-Poissonian
 (super-Poissonian) and $Q=0$ shows the Poissonian statistics.
 For our considered system we have
 \begin{eqnarray}\label{n}
 \langle n \rangle=\sum_{n=\circ}^{\infty}P_{n}\Big[n|A(n,t)|^{2}+(n+1) \left(|B(n+1,t)|^{2}+|C(n+1,t)|^{2} \right)\Big]
 \end{eqnarray}
 and similarly,
 \begin{eqnarray}\label{n2}
 \langle n^{2} \rangle&=&\sum_{n=\circ}^{\infty}P_{n}\Big[n^{2}|A(n,t)|^{2}+(n+1)^{2} \left(|B(n+1,t)|^{2} +|C(n+1,t)|^{2} \right)\Big]
 \end{eqnarray}
 where $ A $, $ B $ and $ C $ have been determined in (\ref{abc}).
 Our presented results in figure 3 show the time evolution of Mandel parameter versus the scaled time $\tau$ for the initial mean number of photons fixed  at $|\alpha|^{2} = 10$. The left (right) plots again correspond to the case $f (n) = 1$ ($f (n) = 1/\sqrt{n}$). Figure 3(a) is plotted for searching the particular effect of intensity-dependent coupling in the absence of Kerr medium and  detuning parameters ($\chi=0,\; \Delta_{2}=\Delta_{3}=0$). The plots of figure 3(a) show that the intensity-dependent coupling causes an increase in the negativity of Mandel parameter in addition to changing its chaotic behaviour between positive and negative values to nearly regular oscillations in the negative region. Figure 3(b) indicates the effect of the Kerr medium ($\chi/\lambda=0.4$) in exact resonance.
 Comparing the left plots of figures of 3(a) and 3(b) shows that the Kerr effect converts some parts of the supper-Poissonian to sub-Poissonian statistics especially at large times, while from the right ones we observe that the Kerr medium causes the regularity of oscillations of the Mandel parameter to be disturbed. In figure 3(c),
 only the effect of detuning parameters is shown ($\Delta_{2}/\lambda=7$ and $\Delta_{3}/\lambda=15$).
 In the left plot of figure 3(c) whereas atom-field coupling does not depend on intensity, we see that the Mandel parameter of the
 field oscillates between negative and positive values. The fact that these parameters take negative values in some intervals of scaled time indicates the nonclassicality feature of states.
 Comparing the left plots of figures of 3(a) and 3(c) shows that the detuning parameters increase the negativity of the Mandel parameter. Also, it is seen that in both of the mentioned figures typical fractional collapses and revivals will appear.
 In general, adding our results displayed in figure 3 (when the right-hand side plots are compared with the left ones), we may easily conclude that
 intensity-dependent coupling has a direct role in revealing this nonclassical property. Indeed, the intensity-dependent
 coupling eliminates the super-Poissonian behaviour of state under consideration such that the observed nonclassical feature (sub-Poissonian behaviour) conserves as time goes on. The right plots of figure 3 show that in the intensity-dependent regime the Mandel parameter is always negative.
 It is useful to indicate that in the absence of the intensity-dependent coupling, the presence of detuning parameters has a significant effect on the maximum amount of negativity of
 the Mandel parameter (strength of nonclassicality of the states), while, in some situations,
 the existence of Kerr medium does not have an expressible influence on this parameter.
 Altogether, by comparing the two models of $\Lambda$- and $V$-type three-level atoms (presented by us and \cite{zait1}, respectively), it can conveniently be understood that the depth of this nonclassicality feature in $\Lambda$-configuration
 is more considerable than $V$-type, i.e, the negativity of the Mandel parameter becomes nearly from $1.5$ to $8$ times greater than
 $V$-type three-level atom.
 %
 %
 \section{Squeezing: normal and higher orders}
 %
 In quantum optics, squeezing phenomenon is describes by decreasing
 the quantum fluctuations in one of the field quadratures with an increase in the  corresponding conjugate quadrature.
 This parameter has been defined in various ways. As some examples, one may refer to first-order and higher-order squeezing.
 We define the following Hermitian operators:
 \begin{eqnarray}\label{vgsp}
 X_{k} = \frac{a^k +a^{\dagger k}}{2}, \hspace{1cm}Y_{k} =
 \frac{a^k -a^{\dagger k}}{2i}, \hspace{1.5cm}
 k=1,2,3,\ldots,
 \end{eqnarray}
 where $k$ indicates the order of squeezing of the radiation field.
 It is worth noticing that higher order squeezing may be generated in higher-order harmonics \cite{zhan}.
 %
 \subsection{Normal (quadrature) squeezing}
 %
 Setting $k=1$ in (\ref{vgsp}), the $X_{1}$ and $Y_{1}$ quadratures obey the commutation relation $[X_{1},Y_{1}]=i/2$.
 Consequently, the uncertainty relation for such operators reads as $\left(\Delta X_{1}\right)^{2}\left(\Delta Y_{1}\right)^{2}\geq 1/16$, where $\langle\Delta Z_{1} \rangle ^{2}=\langle Z_{1}^{2} \rangle-\langle Z_{1} \rangle^{2}$ and $Z_{1}=X_{1}\;\mathrm{or} \; Y_{1}$ and $\Delta X_{1}$ and $\Delta Y_{1}$ are the uncertainties in the quadrature operators $X_{1}$ and $Y_{1}$, respectively.
 A state is squeezed in $X_{1} (Y_{1})$ if $\left(\Delta X_{1}\right)^{2}<0.25\; (\left(\Delta Y_{1}\right)^{2}<0.25)$, or equivalently by defining
 \begin{eqnarray}\label{vsqdx1}
 S_{X}^{(1)}=4 \left(\Delta X_{1}\right)^{2} -1, \hspace{1cm} S_{Y}^{(1)}=4 \left(\Delta Y_{1}\right)^{2} -1,
 \end{eqnarray}
 squeezing occurs in $X_{1} (Y_{1})$ component respectively if $-1<S_{X}^{(1)}<0 \;(-1<S_{Y}^{(1)}<0) $. So the parameters in (\ref{vsqdx1}) sometimes have been called the normalized squeezing parameters.
 These parameters can be rewritten as
 \begin{eqnarray}\label{vsqx1}
 S_{X}^{(1)}&=&2\langle a^{\dag} a \rangle + \langle a^{2} \rangle + \langle a^{\dag 2} \rangle-\left( \langle a \rangle + \langle a^{\dag} \rangle \right)^{2}, \nonumber\\
 S_{Y}^{(1)}&=&2\langle a^{\dag} a \rangle - \langle a^{2} \rangle - \langle a^{\dag 2} \rangle+\left( \langle a \rangle - \langle a^{\dag} \rangle \right)^{2},
 \end{eqnarray}
 where $\langle a^{\dag}a \rangle$ has been given by relation (\ref {n}) and the following general relation can straightforwardly be obtained
 \begin{eqnarray}\label{ar}
 \langle a^{r}\rangle&=&\sum_{n=\circ}^{\infty}q_{n}^{\star}q_{n+r}\Bigg(\sqrt{\frac{(n+r)!}{n!}} A^{\star}(n,t)A(n+r,t) \nonumber \\
 &+& \sqrt{\frac{(n+r+1)!}{(n+1)!}}\left[ B^{\star}(n+1,t)B(n+1+r,t) \right. \nonumber \\
 &+&  \left. C^{\star}(n+1,t)C(n+1+r,t)\right]\Bigg).
 \end{eqnarray}
 Notice that $\langle a^{r} \rangle ^{\star}=\langle a^{\dag\;^{r}}
 \rangle$. Figure 4 describes the first-order squeezing in $X$
 quadrature in terms of scaled time for different chosen parameters
 as before. From the left plot of figure 4(a) with constant coupling,
 in the absence of Kerr medium and in the resonance condition, it is
 seen that the state of the system does not possess squeezing, while
 with the same parameters, if only intensity-dependent coupling
 arrives in the interaction process (right plot of figure 4(a)),
 squeezing will be  seen in the state of the system in all times. It is
 also shown in figure 4(b) that in exact resonance condition and in the presence
 of Kerr medium with either constant or intensity-dependent
 coupling, the results are approximately the same (with no squeezing
 effect). So we may conclude that, while the intensity-dependent
 coupling has a direct role in revealing the squeezing phenomenon,
 the Kerr effect prevents the observation of squeezing of the state of the system.\\
 Finally, figure 4(c) indicates that in the absence of Kerr effect and in non-resonance condition,
 in both cases (constant and intensity-dependent coupling), the results are qualitatively similar to figure 4(a).
 So it seems that detuning parameters do not have a serious effect on the amount of squeezing of the field quadratures in comparison with the intensity-dependent coupling effect.
 By adding the above numerical results of normal squeezing and comparing with similar figures for the $V$-type three-level atom,
 one can see that some of them (left plots of figures 4(a), 4(c) and figure 4(b)) are very similar to their corresponding graphs of \cite{zait1} (both sets are empty from considerable quadrature squeezing).
 Also, the right plot of figure 4(a) is similar, with the important difference that the depth of normal squeezing becomes four times greater in $\Lambda$-type than the $V$-type atom.
 Meantime, comparing the right plot of figure 4(c) with the corresponding one in $V$-type, one observes that in addition to the fact that
 in $\Lambda$-type the system is always quadrature squeezed, the depth of squeezing becomes $10^{3}$ times greater than the $V$-type.
 %
 %
 \subsection{Amplitude-squared squeezing}
 %
 Setting $k=2$ in (\ref{vgsp}), the obtained operators satisfy the commutation relation $\left[X_{2},Y_{2}\right]=i (2n+1)$,
 which obey the uncertainty relation $\left(\Delta X_{2}\right)^{2}\left(\Delta Y_{2}\right)^{2}\geq|\langle n+\frac{1}{2}\rangle|^{2}$.
 Second-order squeezing occurs in $X_{2}\;(Y_{2})$ if $\left( \Delta X_{2} \right)^{2}<|\langle n+\frac{1}{2}\rangle|$($\left( \Delta Y_{2} \right)^{2}<|\langle n+\frac{1}{2}\rangle|$),
 or equivalently, if the following parameters
 \begin{eqnarray}\label{vsqdx2}
 S_{X}^{(2)}=\frac{ \left( \Delta X_{2} \right)^{2}}{|\langle n+\frac{1}{2}\rangle|}-1, \hspace{1cm}  S_{Y}^{(2)}=\frac{ \left(\Delta Y_{2} \right)^{2}}{|\langle n+\frac{1}{2}\rangle|}-1
 \end{eqnarray}
 satisfy the inequality $-1<S_{X}^{(2)}<0\;(-1<S_{Y}^{(2)}<0)$. The latter relations can be rewritten as
 \begin{eqnarray}\label{vsqx2}
 S_{X}^{(2)}&=&\frac{\langle a^{4} \rangle + \langle a^{\dag 4} \rangle + 2 \langle n^{2} \rangle - 2\langle n \rangle - \left( \langle a^{2} \rangle + \langle a^{\dag 2} \rangle \right)^{2}}{4 \langle n \rangle +2}, \nonumber\\
 S_{Y}^{(2)}&=&\frac{2\langle n^{2} \rangle - 2 \langle n \rangle -  \langle a^{4} \rangle - \langle a^{\dag 4} \rangle + \left( \langle a^{2} \rangle - \langle a^{\dag 2} \rangle \right)^{2}}{4 \langle n \rangle +2},
 \end{eqnarray}
 where all of the required quantities in (\ref{vsqx2}) may be
 obtained from (\ref{n}), (\ref{n2}) and (\ref{ar}). Now, we are
 ready to investigate the $X$ component of second-order squeezing.
 Figure 5 shows that amplitude-squared squeezing in  $X$ component for the chosen
 parameters has the same temporal behaviour as normal squeezing. In
 the left plot of figure 5(a), where Kerr effect and intensity-dependent coupling are both absent and in the resonance condition, the
 state of the system does not exhibit squeezing. However, including the
 intensity-dependent coupling causes the state of the system to be
 squeezed in all times (see the right plot of figure 5(a)). Both plots of figure 5(b)
 show the effect of Kerr medium, where it is seen that this effect prevents the
 amplitude-squared squeezing even in the presence of the intensity-dependent coupling. Figure 5(c) has the same temporal
 behaviour as in figure 5(a), which shows the ignorable effect of detuning
 parameters. It seems that this observation is similar to our output results for the normal squeezing discussed at the end of the previous subsection.
 Meanwhile, it is convenient to declare that by comparing both models of $\Lambda$- and $V$-type three-level atoms, where the detuning parameters are nonzero and $f(n)=1/\sqrt{n}$, the $\Lambda$-configuration is always amplitude-squared squeezed, similar
 to the corresponding situation for normal squeezing, while no amplitude-squared squeezing is reported
 in $V$-type \cite{zait1} for a similar situation.
 %
 \subsection{Amplitude-cubed squeezing}
 %
 Putting $k=3$ in (\ref{vgsp}), the operators of third-order squeezing are obtained, which satisfy the commutation relation,
 $\left[X_{3},Y_{3}\right]=\frac{i}{2}\left(9n^{2}+9n+6\right)$.
 The corresponding uncertainty relation reads $\left(\Delta X_{3}\right)^{2}\left(\Delta Y_{3}\right)^{2}\geq \frac{1}{16} |\langle 9n^{2}+9n+6 \rangle|^{2}$.
 Third-order squeezing occurs in $X_{3}\;(Y_{3})$ if $\left( \Delta X_{3} \right)^{2}<\frac{1}{4}|\langle 9n^{2}+9n+6 \rangle|
 \; (\left( \Delta Y_{3} \right)^{2}<\frac{1}{4}|\langle 9n^{2}+9n+6 \rangle|)$, or equivalently, if the parameters
 \begin{eqnarray}\label{vsqdx3}
 S_{X}^{(3)}=\frac{4 \left( \Delta X_{3} \right)^{2}}{|\langle 9n^{2}+9n+6 \rangle|}-1, \hspace{1cm} S_{Y}^{(3)}=\frac{4 \left( \Delta Y_{3} \right)^{2}}{|\langle 9n^{2}+9n+6 \rangle|}-1
 \end{eqnarray}
 satisfy the inequality $-1<S_{X}^{(2)}<0\;(-1<S_{Y}^{(2)}<0)$.
 Consequently, the latter relations can be rewritten as
 \begin{eqnarray}\label{vsqx3}
 S_{X}^{(3)}&=&\frac{\langle a^{6} \rangle + \langle a^{\dag 6} \rangle + 2 \langle n^{3} \rangle +4\langle n \rangle - 6\langle n^{2} \rangle - \left( \langle a^{3} \rangle + \langle a^{\dag 3} \rangle \right)^{2}}{9\langle n^{2} \rangle +9\langle n \rangle +6}, \nonumber\\
 S_{Y}^{(3)}&=&\frac{2\langle n^{3} \rangle - 6 \langle n^{2} \rangle + 4 \langle n \rangle - \langle a^{6} \rangle - \langle a^{\dag 6} \rangle + \left( \langle a^{3} \rangle - \langle a^{\dag 3} \rangle \right)^{2}}{9\langle n^{2} \rangle +9\langle n \rangle +6}.
 \end{eqnarray}
 We have plotted the amplitude-cubed squeezing for different chosen parameters mentioned in figure 6. It is seen from the left plot of figure 6(a)
 that in the absence of Kerr effect and in resonance condition, the state of the system does not have amplitude-cubed squeezing property.
 However, after entering the intensity-dependent coupling and by maintaining the other conditions, the state will be amplitude-cubed squeezed
 in all times (the right plot of figure 6(a)). Figure 6(b) indicates the effect of Kerr medium, from which one can see that the squeezing does not occur unless in some sharp finite points of time in the presence of intensity-dependent coupling.
 We again found that the existence of the Kerr medium causes the reduction of the amplitude-cubed squeezing features of the state of the system even in the presence of the intensity-dependent coupling.
 At last, figure 6(c) describes  the effect of detuning parameters from which nearly similar behaviour to figure 6(a) is observed.
 For the special case where we have the non-resonance condition and intensity-dependent coupling, the behaviour of amplitude-cubed squeezing in both $\Lambda$- and $V$-type models are obviously different. It is found that, only $\Lambda$-type configuration exhibits the amplitude-cubed squeezing property. The right plots of figures 4-6 indicate that
 with the increase of the order of squeezing, the amount of squeezing is decreased.\\
 We conclude this section with emphasizing that, while generally the atom-field intensity-dependent coupling has an obvious role in revealing different orders of the squeezing of the state of the system, the effect of detuning parameters is weaker (especially, when $f(n)=1$, i.e. no intensity-dependent coupling exists)
 and the Kerr medium attenuates or even sometimes prevents them from being observed.
 %
 %
 \section{Quasiprobability $Q$-distribution function}
 %
 We will now calculate the  Quasiprobability $Q$-distribution function \cite{zubairy}, numerically.
 For the atom-field system under consideration, one can obtain
 \begin{eqnarray}\label{q3}
 Q(\alpha ,t)&=&\frac{1}{\pi}|\langle \alpha |\psi \rangle|^{2}=\frac{1}{\pi}e^{-|\alpha|^{2}}\sum_{n=0}^{\infty}\frac{|\alpha|^{2n}}{n!} P_{n} \nonumber \\
 &\times&\Big[|A(n,t)|^{2}
 + \frac{|\alpha|^{2}}{n+1}\Big(|B(n+1,t)|^{2}+|C(n+1,t)|^{2}\Big) \Big].
 \end{eqnarray}
 In figure 7, the three-dimensional distribution function and the corresponding contour plots in phase space have been sketched for some
 chosen parameters, where  $x=\mathrm{Re}(\alpha)$ and $y=\mathrm{Im}(\alpha)$ and we assumed $\lambda t=\pi/2$.
 Figure 7(a) corresponds to the exact resonance condition with no Kerr  effect and constant atom-field coupling ($f(n)=1$).
 In figure 7(b), all chosen parameters are the same as in figure 7(a), except that the coupling has been considered to be intensity-dependent ($f(n)=1/\sqrt{n}$).
 By comparing these two figures, we conclude that in the presence of the intensity-dependent coupling, distribution function is well spread and it is seen that the maximum amount of this function is obviously reduced. The influence of Kerr medium on the distribution function in the resonance condition and with constant coupling has been shown in figure 7(c). This figure indicates that quasiprobability distribution function includes a hole, which is caused by the Kerr medium.
 Also, the intensive drop of the maximum amount of this  function can be observed due to the influence of the Kerr effect.
 %
 %
 \section{Summary, discussion and concluding remarks}
 %
 In this paper, we have considered the nonlinear interaction between a $\Lambda$-type three-level atom and a single-mode field
 in a cavity containing a Kerr medium  using the generalized JCM  with intensity-dependent coupling between the atom and the field. Next, after finding the explicit form of the state vector of the atom-field system in a general manner, field entropy, quantum statistics,
 some nonclassical properties and the quasiprobability distribution function of the obtained state have been investigated, numerically.
 In particular, we studied the effects of $` `$intensity-dependent coupling$"$ (by considering the nonlinearity function as in \cite{zait1} $f(n)=1/{\sqrt n}$),
 $` `$Kerr medium$"$ and $` `$detuning parameters$"$ on the mentioned physical quantities, individually. We would like to emphasize the generality of our obtained formalism in the sense that it may be used for any physical system, either any nonlinear oscillator algebra with arbitrary $f(n)$, or any solvable quantum system with known $e_{n}$, using the relation $e_{n}=nf^{2}(n)$ \cite{honarasa}.
 In this section, after presenting a summary of our extracted results, we briefly compare them with those of $V$-type atoms which have been reported in \cite{zait1}.
 Firstly, focusing our attention to the obtained numerical results for $\Lambda$-type atoms, one can see that
 generally intensity-dependent coupling reduces the maximum (and also minimum) amounts of field entropies, unless the case which the system is in resonance condition and the Kerr medium is not present.
 Also, it causes a regular oscillatory behaviour in the field entropy, except where the Kerr medium exists (right plot of figure 2(b)).\\
 Our further calculations on the nonclassical properties such as sub-Poissonian statistics and different orders of squeezing  (consist normal or quadrature, amplitude-squared and amplitude-cubed squeezing) indicated that, if atom-field coupling depends on the intensity of the field,  the mentioned nonclassicality signs will be generally changed, except in the cases in which the Kerr medium is present.
 In more detail, we may imply that intensity-dependent coupling causes the Mandel parameter to take negative values at all times for all considered cases,
 although its depth has been reduced. This means that intensity-dependent coupling converts all parts of the
 figures with supper-Poissonian behaviour to sub-Poissonian statistics. This stated result is true in the presence of Kerr medium and the detuning parameters, too.
 This effect ($f(n)=1/{\sqrt n}$) for squeezing behaviour seems to be somewhat different. Indeed, unless the case in which the Kerr medium exists,
 intensity-dependent coupling reveals all orders of squeezing at all times with a more considerable depth.
 In this way, the major effect of intensity-dependent coupling on the range and strength of these particular nonclassicality features of the state vector
 of the system under consideration is well illustrated. So it is not far from reality if we say that the intensity-dependent coupling  plays an important
 role in ascending the latter quoted nonclassical properties of the system, and the Kerr medium prevent them from being observed.\\
 As is observed from the left plots of Figures 4-6 ($f(n)=1$), no
 squeezing is seen, exept in some cases and in a very short time intervals.
 On the other hands, from the right plots of Figures 4-6 ($f(n)=1/\sqrt n$), it is obvious that squeezing of all orders occurrs and their strengths quantitatively
 decrease with increasing order of squeezing.
 Indeed, squeezing of all orders (consist normal or quadrature, amplitude-squared and amplitude-cubed) is seen at all times, except in the presence of Kerr medium.\\
 Now, if one compares our presented computational results with the similar situations (conditions)
 for three-level atom in $V$-configuration presented in \cite{zait1},
 it is observed that the investigated physical properties in
 $\Lambda$-configuration are more visible than $V$-type (of course for the chosen parameters, although they have selected randomly).
 For instance, by comparing the time evolution of the field entropy for
 $\Lambda$-type (all graphs displayed by us in figure 2) with the same
 quantity for $V$-type three-level atom in \cite{zait1}, it is seen
 that the maxima of entropies are considerably increased in our considered
 $\Lambda$-configuration. Specifically, for the case
 $\chi=\Delta_{2}=\Delta_{3}=0$ with $f(n)=1/\sqrt{n}$, one can see
 that the maxima in the entropy are highly increased by the order of
 approximately $\simeq 10^{2}$ times. Continuing the comparison to the
 Mandel parameter, it will be seen that the sub-Poissonian
 criterion (negativity of $Q$-parameter ) of $\Lambda$-type is enhanced quantitatively from the
 order of $1.5$ up to $8$ times relative to $V$-type configuration.
 Approximately, the same results may be illustrated for the
 squeezing parameters. But, before arriving at this special discussion
 in detail, it is worth mentioning that our definitions of different
 orders of squeezing were chosen to be normalized, i.e. the amount
 of squeezing for any order lies  between the values $-1$ and $0$ (it has the
 lower bound of $-1$). So, to have a better comparison, we had to
 repeat the numerical calculations of higher order squeezing
 according to our definition, for $V$-type atoms (since the
 definition used in \cite{zait1} for these quantities
 is not like ours). The obtained results enabled us to compare
 the squeezing features for $\Lambda$- and $V$-type configurations.
 For instance, as an illustrative example one
 may refer to figure 4(c). From this figure it is observed that for
 $\Lambda$-configuration for intensity-dependent coupling and with nonzero detuning parameters, the state
 of the system is always normally squeezed, but in
 $V$-configuration using similar parameters, the state of the
 system does not have squeezing at all.
 The temporal behaviour of higher order squeezing for
 $V$-type and $\Lambda$-type is qualitatively similar to the normal squeezing, except the cases in which the effects of detuning parameters and
 intensity-dependent coupling have been simultaneously remarked.
 In this direction, consider figures 4(c), 5(c) and 6(c),  i.e. the cases in which $\chi=0, \Delta_{2}/\lambda=7$
 and $\Delta_{3}/\lambda=15$ with $f(n)=1/\sqrt{n}$. In these cases,
 while for $V$-type atoms amplitude-squared and amplitude-cubed squeezing are not reported \cite{zait1}, in the $\Lambda$-type configuration, these nonclassicality
 indicators are always seen as time goes on.
 After all, we observe that in the  $\Lambda$-type configuration
 the squeezing properties are  considerably enhanced relative to
 $V$-type. For example, from figure (4), it is seen that in the presence of intensity-dependent coupling, the depth of squeezing feature in $\Lambda$-configuration
 is almost $4$ (in the resonance condition) to $10^3$ (in the presence of detuning parameters) times greater than similar situations for $V$-type atoms.
It is worth mentioning that, recently atomic squeezing (as another nonclassical effect) for all types of three-level atoms consist of ladder-,
$V$- and $\Lambda$-configurations interacting with a radiation field has been studied by Civitarese {\it et al} \cite{atomsqu}.
According to their results, atomic squeezing becomes evident in both ladder and $\Lambda$ schemes of three-level atoms.
In addition, they found that, regardless of the choice of the coupling constants and the number of atoms and photons,
spin squeezing does not appear so clearly in $V$-type three-level atoms. Our results, in a sense, are further evidences
on their work, i.e. the nonclassical properties (such as the degree of
entanglement between atom and field, sub-Poissonian statistics and different orders of squeezing) are also not so observable in $V$-type,
in comparison with $\Lambda$-type three-level atoms.
Therefore, even though Zait \cite{zait1} (and so we) has chosen the parameters randomly,
altogether, since our results are based on the numerical calculations, and the effective parameters which enter the interaction are various,
as we repeatedly emphasized, the strongness of each nonclassicality criterion for the $\Lambda$-type atoms depends on the chosen parameters.
\\
 At last, it is appropriate to dip into the influences
 of the Kerr medium and intensity-dependent coupling on the behaviour of the quasiprobability distribution function (figure (7)).
 It appears that intensity-dependent coupling spreads this function while the Kerr medium causes the creation of a hole on the $Q$-distribution function.
 Also, it is seen that both of the latter effects can reduce the maximum amount of
 $Q$-function.\\
 At the end of this paper, we mention that this study can be accomplished by considering a two-mode field for both configurations of $\Lambda$- and $V$-type
 three-level atoms and different initial states of the atom-field system. These works are in preparation and will be submitted in near future.

 \begin{flushleft}
 {\bf  Acknowledgments}\\
 \end{flushleft}
 The authors would like to thank the referees who have made valuable and careful comments about some
 points relating to our manuscript, which improved the paper considerably. They are thankful to Dr M R Hooshmandasl for his useful assistance in the numerical results.
 One of the authors (MJF) would like to thank Dr G R Honarasa for his intuitive comments and valuable discussions during preparation.
 Also, he is grateful to M R Faghihi for his kind help during this work.
 \vspace {2 cm}
 \end{document}